\documentclass[usegraphicx,usenatbib]{mn2e}

\title[The SFR-density relation up to $z\sim3$]{ The relationship between star formation rates, local density and stellar mass up to $z\sim3$ in the GOODS NICMOS Survey}

\author[Gr\" utzbauch et al.]{R. Gr\"utzbauch$^{1}$\thanks{email: ruth.grutzbauch@nottingham.ac.uk}, C. J. Conselice$^{1}$, A. E. Bauer$^{2}$, A. F. L. Bluck$^{3}$, \newauthor R. W. Chuter$^{1}$, F. Buitrago$^{1}$, A. Mortlock$^{1}$, T. Weinzirl$^{4}$, S. Jogee$^{4}$\\
$^{1}$ School of Physics and Astronomy, University of Nottingham, UK \\
$^{2}$ AAO, Sydney, Australia \\
$^{3}$ Gemini Observatory, Hilo, Hawaii, USA \\
$^{4}$ Department of Astronomy, University of Texas, USA 
}

\begin{document}

\date{Accepted -- . Received --}

\maketitle

\begin{abstract}

We investigate the relation between star formation rates and local galaxy environment for a stellar mass selected galaxy sample in the redshift range $1.5 \leq z \leq 3$. We use near-infra-red imaging from an extremely deep Hubble Space Telescope survey, the GOODS-NICMOS Survey (GNS) to measure local galaxy densities based on the nearest neighbour approach, while star-formation rates are estimated from rest-frame UV-fluxes. Due to our imaging depth we can examine galaxies down to a colour-independent stellar mass completeness limit of $\log~M_\ast = 9.5$ M$_\odot$ at $z\sim3$.
We find a strong dependence of star formation activity on galaxy stellar mass over the whole redshift range, which does not depend on local environment. The average star formation rates are largely independent of local environment apart from in the highest relative over-densities. Galaxies in over-densities of a factor of $>5$ have on average lower star formation rates by a factor of $2-3$, but only up to redshifts of $z\sim2$. We do not see any evidence for AGN activity influencing these relations. We also investigate the influence of the very local environment on star-formation activity by counting neighbours within 30~kpc radius. This shows that galaxies with two or more close neighbours have on average significantly lower star formation rates as well as lower specific star formation rates up to $z\sim 2.5$. We suggest that this might be due to star formation quenching induced by galaxy merging processes.

\end{abstract}

\section{Introduction}

Observational studies have shown that up to half of the currently existing stellar mass was already in place by $z\sim1$ \citep{Bri00,Dro04,Bun06,Per08,Mor11}. Similarly, the peak of cosmic star formation activity has been observed to occur at a redshift of $z>1$ \citep[see e.g.,][]{Mad96,Cim04,Jun05}, and additionally seems to depend on galaxy mass: the most massive galaxies are among the first galaxies to form \citep[e.g.,][]{Tho05,Cle06,Cim06} and are roughly in place by a redshift of $z \sim 1$, where their number densities reach similar values as in the local Universe \citep[e.g.,][]{Con07}. These early massive galaxies at $z>1.5$ are found to have on average red rest-frame colours as expected for a maximally old stellar population at these very early epochs \citep{Gru11b}, however, \citet{Bau11} show that 50 - 80\% of these massive red galaxies might harbour dusty star formation \citep{Bau11}.

It is then clear that the main epoch of galaxy formation occurs at $z>1$, but it is still unclear which processes trigger or suppress star formation, and what are the most important factors in determining galaxy properties. Is it primarily galaxy mass (i.e. a galaxy's gravitational potential), is it galaxy merging on local scales or is it the large scale environment that drives galaxy evolution?

Various processes connected to a galaxy's environment are expected to trigger or suppress star formation and influence galaxy assembly \citep{GG72,Lar80,Mih94,Moo96}. The efficiency of these processes depends on the local galaxy density (e.g., for minor and major merging or harassment) or the density of the intra-cluster medium (e.g., for ram-pressure stripping). We might then expect to find a correlation between local density and star formation rate, as it is indeed found in the local Universe \citep[e.g.,][]{Lew02,Bal04} and up to $z\sim1$ \citep[e.g.,][]{Elb07,Coo08,Pat09,Sob10}, however with varying conclusions. At lower redshifts star formation rates of galaxies in clusters are found to decrease both with smaller cluster-centric distance and as a function of local projected galaxy density \citep[e.g.,][and references therein]{Pog99,Lew02}. \citet{Bal04} show that the correlation between local density and star formation rate extends beyond the dense cluster environment to galaxies in the field. They argue that short-term local processes like galaxy merging at high redshift causes this dependence. At $z\sim1$, some authors report a turn-around of the local SFR-density relation \citep{Elb07,Coo08}, corresponding to enhanced star formation activity in areas of high local density. Others \citep[e.g.,][]{Pog08,Pat09} find that star formation is suppressed in the highest density areas. This effect is observed in the cores of the most massive galaxy clusters as early as $z\sim1.4$ \citep{Lid08,Bau11}, while in less massive clusters at similar redshifts star formation in the cluster core was found to be enhanced \citep{Hay10,Tra10}.

Recently, \citet{Sob10} presented a study of SFRs over a wide range of environments at $z\sim1$ and found that the median SFR and star-forming fraction increases with local density up to a critical surface density of $\Sigma =$ 10-30 Mpc$^{-2}$ and decreases at higher densities, possibly partly reconciling the contradictory results of the above studies. However, lower SFRs and redder galaxy colours were also observed in general field samples outside the dense cores of galaxy clusters up to $z\sim1.8$ \citep{Qua11,Chu11}. The controversy of the role of environmental processes in influencing the build-up of galaxy stellar mass at $z>1$ still remains. 

So when and where does the relation between star formation and environment emerge? To witness the transformation from heavily star-forming to passively evolving galaxies we have to move to the crucial redshift domain at $z\sim1.5$ and above. In this epoch most of the stellar mass is built up and the formation of massive galaxies is still on-going \cite[e.g.,][]{Fon06,Mor11}. To investigate the influence of galaxy environment on this process, a very deep, high spatial resolution survey, that allows for the detection of faint satellites, is crucial. In this study we utilise data from the GOODS NICMOS Survey (GNS), an extremely deep, near infrared Hubble Space Telescope survey, reaching unprecedented stellar mass completeness of $\log~M_\ast= 10^{9.5} M_\odot$ up to $z=3$. Using the same sample, \citet{Gru11b} found that the colour-density relation is possibly reversed at $z\sim 1.5-2$ and is not detectable at $z>2$. In the present study we investigate the behaviour of star formation rates as a function of local density as well as the possible differences between the SFR-density and colour-density relation introduced by the presence of strong dust extinction. Using a stellar mass limited sample also allows us to investigate the role of stellar mass relative to that of local environment without being biased towards star-forming galaxies, which is often the case for optically selected samples.

We present this data and the measurements of stellar masses, star formation rates and local densities in Section~\ref{DandA}. The results are shown in Section~\ref{Results}, while we discuss and summarise our findings in Section~\ref{summary}. Throughout the paper we assume the standard $\Lambda$CDM cosmology, a flat universe with $\Omega_\Lambda = 0.73$,  $\Omega_M = 0.27$ and a Hubble constant of $H_0 = 72 $ km s$^{-1}$ Mpc$^{-1}$.

\section{Data and analysis}\label{DandA}

In this section we describe the survey we use in this study, the GOODS NICMOS Survey (GNS), as well as the measurements of photometric redshifts, stellar masses, rest-frame colours, local densities and star formation rates.

\subsection{The GNS data}\label{data}

The data used in this study is obtained with the GOODS NICMOS Survey (GNS). The GNS is a 180 orbit Hubble Space Telescope survey consisting of 60 single pointings with the NICMOS-3 near-infrared camera, with an imaging depth of three orbits per pointing. 

The pointings were optimised to contain the maximum number of massive galaxies ($M_\ast > 10^{11}~M_\odot$) in the redshift range $1.7 < z < 3$, identified in the two GOODS fields by their optical-to-infrared colours \citep{Con10}. The survey covers a total area of about 45 arcmin$^2$ with a spatial resolution of $\sim$0.1 arcsec/pixel, corresponding to $\sim$0.9~kpc at the redshift range we use here ($1.5 \leq z \leq 3$). The target selection, survey characteristics and data reduction are fully described in \citet{Con10}. For first science results from the survey see e.g., \citet{Bui08}, \citet{Blu09}, \citet{Gru11b} and \citet{Bau11}.

The GNS has a 5$\sigma$ limiting magnitude of $H_{AB}$ = 26.8, which is significantly deeper than ground based near-infrared imaging of the GOODS fields done with ISAAC on the VLT, reaching a 5$\sigma$ depth of $H_{AB} = 24.5$ \citep{Ret10}. Sources were extracted from the NICMOS $H$-band image and matched to the optical HST-ACS bands $B$, $V$, $i$ and $z$, which is available down to a limiting magnitude of $B_{AB} = 28.2$. The matching is done within a radius of $2 ^{\prime\prime}$, however the mean separation between optical and $H$-band coordinates is much better with $\sim 0.28 \pm 0.4 ^{\prime\prime}$, roughly corresponding to the NICMOS resolution \citep[see also][]{Bau11}.

The resulting $H$-band selected photometric catalogue covering the bands $BVizH$ comprises about 8300 galaxies and is used to compute photometric redshifts, rest-frame colours and stellar masses as described in the following sections \citep[see also ][ for more details]{Con10}.

\subsection{Photometric Redshifts}\label{photoz}

Photometric redshifts were obtained by fitting template spectra to the $BVizH$ photometric data points using the HYPERZ code \citep{Bol00}. The method is described in more detail in \citet{Gru11b}. The synthetic spectra used by HYPERZ are constructed with the Bruzual \& Charlot evolutionary code \citep{Bru93} representing roughly the different morphological types of galaxies found in the local universe. We use five template spectra corresponding to the spectral types of E, Sa, Sc and Im as well as a single burst scenario. The reddening law is taken from \citet{Cal00}. HYPERZ computes the most likely redshift solution in the parameter space of age, metallicity and reddening. The best fit redshift and corresponding probability are then output together with the best fit parameters of spectral type, age, metallicity, $A_V$ and secondary solutions.

To assess the reliability of our photometric redshifts we compare them to available spectroscopic redshifts in the GOODS fields \citep{Bar08,Wuy08}. We matched the two catalogues to our photometric catalogue with a matching radius of $2 ^{\prime\prime}$, obtaining 906 secure spectroscopic redshifts.  The reliability of photometric redshift measures is usually defined as $\Delta z/(1+z) \equiv (z_{spec}- z_{phot})/(1+z_{spec})$. In the following we compare the median offset from the one-to-one relation between photometric and spectroscopic redshifts, $\langle \Delta z/(1+z) \rangle$, and the RMS scatter around this relation, $\sigma_{\Delta z/(1+z)}$. 
For the whole sample over the full redshift range we obtain a median offset of $\langle \Delta z/(1+z) \rangle = 0.011$ and a scatter of $\sigma_{\Delta z /(1+z)} = 0.061$. We then investigate the performance of HYPERZ at different redshifts, at low redshift ($z<1.5$) and in the redshift range of $1.5 \leq z \leq 3$, which is the redshift range of the galaxy sample we use in this study. 
For the high redshift sample we obtain an average offset $\langle \Delta z/(1+z) \rangle = 0.06$ and a RMS of $\sigma_{\Delta z /(1+z)} = 0.10$, with a fraction of catastrophic outliers of $20\%$, where catastrophic outliers are defined as galaxies with $|\Delta z/(1+z)| > 0.3$, which corresponds to $\sim$ 3 times the RMS scatter. Galaxies below $z=1.5$ show a slightly lower, but still comparable scatter of $\sigma_{\Delta z /(1+z)} = 0.08$, however the outlier fraction decreases dramatically to only $\sim 2\%$.
These are the values that are used in the Monte Carlo Simulations described in Section~\ref{MC} to account for the effect of the photometric redshift errors on the local density and SFR estimates.

\subsection{Stellar masses and rest frame colours}\label{masses}

The determination of stellar masses for our sample is fully described in \citet{Con10} and \citet{Mor11}. The stellar masses and rest-frame colours we use are measured by multicolour stellar population fitting techniques, based on the same catalogue used for the photometric redshift measurements. Spectroscopic redshifts are used if available. A large set of synthetic spectral energy distributions (SEDs) is constructed from the stellar population models of \citet[][BC03]{Bru03}, assuming a Salpeter initial mass function. The star formation history is characterised by an exponentially declining model with various ages, metallicities and dust extinctions. The model SEDs are then fit to the observed photometric datapoints of each galaxy using a Bayesian approach. For each galaxy a likelihood distribution for the stellar mass, age, and absolute magnitude at all star formation histories is computed. The peak of the likelihood distribution is then adopted as the galaxy's stellar mass and absolute $U$ and $B$-band magnitudes, while the uncertainty of these values is given by the width of the distribution. We chose to compute rest-frame $(U-B)$ colours, since the wavelength range of the $U$ and $B$ bands is covered best by the observed optical and H bands.

While parameters such as age, e-folding time and metallicity are not accurately fit due to various degeneracies, the stellar masses and colours are robust. From the width of the probability distribution we determine typical errors for our stellar masses, which are mainly due to uncertainties in the template fitting and photometric errors.
There are additional uncertainties from the choice of the IMF, which are not taken into account here. We obtain a  total random error of our stellar masses of $\sim$0.3 dex, roughly a factor of two. A more detailed discussion of the stellar mass uncertainties can be found in \citet{Con10}.

We do not use any photometric data redder than the NICMOS $H$-band in our measurements of stellar masses.   The reason for this is that essentially there is no data redder than the $H-$band which has the same fidelity and depth as the $BVizH$ bands we use in this paper.  The K-band data available from the ground is nowhere near as deep as the $H-$band NICMOS data.  While we have IRAC data for our sources, we do not use these data due to the PSF issues and contamination from neighbouring galaxies.    Furthermore, the rest-frame $B$-band gives us a good anchor for measuring stellar masses, as is shown by e.g., \citet{BdJ01} at lower redshifts, and \citet{Bun06} for higher redshifts.

\subsection{Completeness limits}\label{completeness}

We compute the expected stellar mass completeness limits from the 5$\sigma$ magnitude limit of the GNS ($H_{AB}$ = 26.8) and the mass-to-light ratios (M/L) of simple stellar populations (SSP). For more details of this procedure we refer the reader to \citet{Con10} and \citet{Mor11}. The limiting stellar mass for a galaxy with a maximally old stellar population at $z = 3$ is $M_\ast = 10^{9.2} M_\odot$.
The influence of the presence of dust on the stellar mass completeness limit is investigated in \citet{Bau11}. They find that a very high dust extinction of $A_V > 3$ magnitudes is necessary to remove low-mass heavily dust-obscured star-forming galaxies from our sample. It is therefore unlikely that our sample is significantly biased against dusty star-forming objects.
To make sure that our sample is not biased towards blue galaxies at high $z$ we use a conservative stellar mass cut of $\log M_\ast = 9.5$ in the following. The colour dependent completeness limits and mass functions of red and blue galaxies in the GNS are investigated in more detail by \citet{Mor11}. 

In this study we will focus on the redshift range of $1.5 < z < 3$, which provides a co-moving survey volume of $\sim2.3\times10^5$ Mpc$^3$, minimising the effects of cosmic variance. The cosmic variance associated with a certain co-moving volume depends on the clustering strength of the objects and can be roughly estimated from the average galaxy number density and the expected variance of dark matter halos at a certain redshift \citep{Som04}. For the co-moving volume of the GNS within the redshift range of $1.5 < z < 3$, and at the average stellar mass of galaxies in the survey, we expect the influence of cosmic variance to be less than 10\%.
The final galaxy catalogue we use in the following comprises 1289 galaxies down to a stellar mass of $\log M_\ast = 9.5$ within the redshift range of $1.5 < z < 3$.

\subsection{Local densities}\label{densities}

The local densities we use here are measured by \citet{Gru11b} using the same sample to perform a study of galaxy colours and blue fractions in different environments. Two different approaches are considered in the above study, (1) the aperture density, based on galaxy counts in a fixed physical aperture and (2) the nearest neighbour density, based on the distance to the 3$^{rd}$, 5$^{th}$ and 7$^{th}$ nearest neighbour. Galaxies down to the completeness limit of $\log M_\ast = 9.5$ are taken into account here. For both methods a redshift interval of $\Delta z = \pm 0.25$ is used to minimise the contamination of our sample with foreground or background objects. \citet{Gru11b} show that the 3$^{rd}$ nearest neighbour density is best suited to identify extremes in the local density distribution, which is why we will use it as a local density estimator in this study.

We give an overview of the method in the following. First, the distance to the third nearest galaxy, $D_3$, within the redshift interval of $\Delta z = \pm0.25$ is computed for each galaxy. Secondly, we have to account for edge effects, which are a major issue due to the design of the GNS, since it does not have a continuous survey area. The area containing the three nearest neighbours might not be fully covered for all galaxies, especially for galaxies close to the edges of isolated, non-overlapping pointings, since the field of view of a single pointing covers about 500 kpc at $z > 1.5$.  To properly account for the loss of area due to survey edges we approximate the sampled area around each galaxy by the number of image pixels, which are actually covered by the observations, within the area $\pi D_3^2$. Since the number of pixels within $\pi D_3$ ($N_{pixels}(D_3)$) is directly proportional to the covered area, it can be used instead of $\pi D_3^2$ to compute the surface density $\Sigma_3 = 3/N_{pixels}(D_3)$. This gives $\Sigma_3$ in arbitrary units, which is ideal for our purposes, since we are only interested in relative densities and $\Sigma_3$ is normalised by the median value in each redshift slice, $\langle \Sigma_3\rangle_{\Delta z}$. The nearest neighbour density in units of a {\it relative over-density} is then given by:

\begin{equation}
(1+\delta_3) = \frac{\Sigma_3}{ \left\langle \Sigma_3\right\rangle_{\Delta z}}
\end{equation}

\noindent where $\delta_3$ itself is the over-density. In the following we use $\log~(1+\delta_3)$ to distinguish between under-dense ($\log~(1+\delta_3) < 0$) and over-dense ($\log~(1+\delta_3) > 0$) environments.

\subsection{Star formation rates}\label{SFR}

The star formation rates (SFRs) were measured from the rest-frame ultraviolet (UV) luminosities, as fully described in \citet{Bau11}. The rest-frame UV provides one of the most direct measurements of ongoing SFR, since the UV luminosity is directly related to the presence of a young and short-lived stellar population produced by recent star formation. However, UV light is very susceptible to dust extinction and a careful dust-correction has to be applied.  The correction we use here is based on the rest-frame UV slope, i.e. a far-to-near UV colour.
The method is described in \citet{Bau11}, presenting a detailed study the star-formation properties of the same sample we use here. We briefly describe the method in the following.

We determine the SFR$_{UV}$ from the observed optical ACS/$z_{850}$-band flux density (with a 5~$\sigma$ limit of 27.5 in the AB system), which corresponds to the rest-frame UV luminosity around 2800$\mathrm{\AA}$, spanning wavelengths of 2125~-~3400$\mathrm{\AA}$ for $z=1.5-3$ galaxies.  First we derive absolute magnitudes using the {\tt kcorrect} package \cite[v4.2]{Bla07}.  The 2800$\mathrm{\AA}$ luminosity is then converted into a star formation rate assuming a Salpeter IMF and using the \citet{Ken98} law:

\begin{equation} \label{eq:sfr2}
\textrm{SFR}_{\mathrm{UV}}\,(M_\odot \,\textrm{yr}^{-1}) = 1.4 \times10^{-28}\, L_{2800}\,(\textrm{ergs }\,\textrm{s}^{-1}\,\textrm{Hz}^{-1})
\end{equation}

\noindent A dust correction is determined for each galaxies uniquely by fitting model SEDs to the observed galaxy magnitudes from the optical to the mid-infrared, following the procedure described in \citet{Per08}. The best-fitting template is used to obtain synthetic UV luminosities at 1600$\mathrm{\AA}$ and 2800$\mathrm{\AA}$.   The UV-slope $\beta$ is then used to derive the extinction at 2800$\mathrm{\AA}$, $A_{2800}$ following the law calibrated by \citet{Cal00}.

The depth of the $z$-band data from HST-ACS allows us to reach limiting SFRs of 1.5~$M_\odot$yr$^{-1}$ at $z=1.5$ and 5~$M_\odot$yr$^{-1}$ at $z=3$.  These values are determined by using the $z$-band limit plus a one magnitude dust correction to calculate the SFR across the redshift range used in this study.  Of the 1289 galaxies within the redshift range of $1.5 \leq z \leq 3$ and stellar masses above $M_*= 10^{9.5}$$M_\odot$, the majority of galaxies show signs of star formation, with only 1.5\% not detected in the $z$-band.

\subsection{Monte Carlo simulations}\label{MC}

To estimate the reliability of the local density estimates we perform a set of Monte Carlo simulations, using the same method as in \citet{Gru11b}. To asses the effect of the photometric redshift uncertainties on the local densities and star formation rates we randomise the input redshift catalogue according to the $\Delta z / (1+z) $ error obtained by the comparison with available secure spectroscopic redshifts in Section~\ref{photoz}. We use the typical photometric redshift error $\Delta z / (1+z) = 0.10$ in our redshift range of $1.5 \leq z \leq 3$ and assume a Gaussian distribution of errors, where the width of the distribution $\sigma$ corresponds to $\Delta z / (1+z) $. For each galaxy a random value is selected within this distribution, which is then added to the measured photo-z. To account for scattering in and out of the redshift range we include the full GNS sample in the randomisation and compute the redshift uncertainty for each redshift range. We obtain an average $\Delta z / (1+z) = 0.08$ for galaxies with redshifts lower than our range of interest ($z < 1.5$) and an average $\Delta z / (1+z) = 0.1$ for galaxies with redshifts higher than the range we use in this study ($z>3$). We deal with catastrophic outliers separately by randomly adding much larger offsets for a percentage of galaxies, corresponding to the catastrophic outlier fraction at the respective redshift (see Section~\ref{photoz}). Galaxies with $|\Delta z / (1+z)| > 0.3$ are treated as catastrophic outliers here. The offsets are randomly picked from the interval $0.3 < \Delta z / (1+z) < 1$, and added or subtracted from the original redshift. 

\begin{figure*}
\includegraphics[angle=270, width=0.9\textwidth]{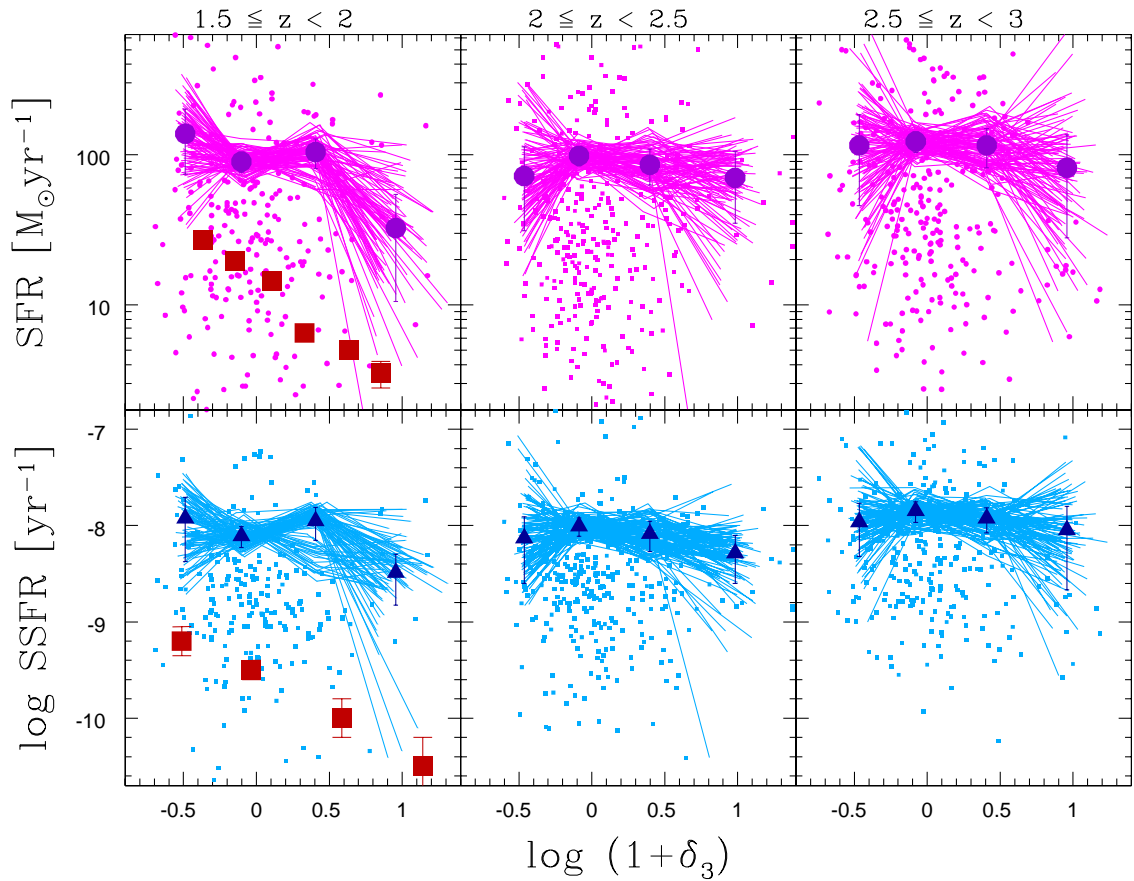}
\caption{Dust corrected star formation rates (SFR, top) and specific star formation rates (SSFR, bottom) as a function of local density $\log ~(1+\delta_3)$. The small symbols are the observed data, while the big symbols represent the average of all Monte-Carlo simulations. Each line is the result of a single Monte-Carlo run. The red squares are lower redshift data from \citet{Lew02} (SFR, upper panel) and \citet{Pat11} (SSFR, lower panel), see text for details. }
\label{SFR_SSFR_SED}
\end{figure*}

The randomisation procedure is repeated 100 times for each galaxy in the sample, resulting in a randomised photometric redshift catalogue, which is then used to recalculate the local densities and star formation rates for each of the 100 Monte Carlo runs. From this we obtain 100 randomised local density estimates for each galaxy in the sample, from which we then compute the average local density uncertainty for the whole sample. The local density error for each individual galaxy is simply the standard deviation of all Monte Carlo runs. The individual errors are then averaged to obtain the average uncertainty of the local density estimator, which is $\Delta \log~(1+\delta_3) = 0.24$. The SFR of each galaxy is recomputed using the change in luminosity distance caused by the change in photometric redshift in each simulation to obtain a new UV luminosity $L_{2800}$, which is then used in equation~\ref{eq:sfr2}, as above.

\begin{figure*}
\includegraphics[angle=270, width=0.9\textwidth]{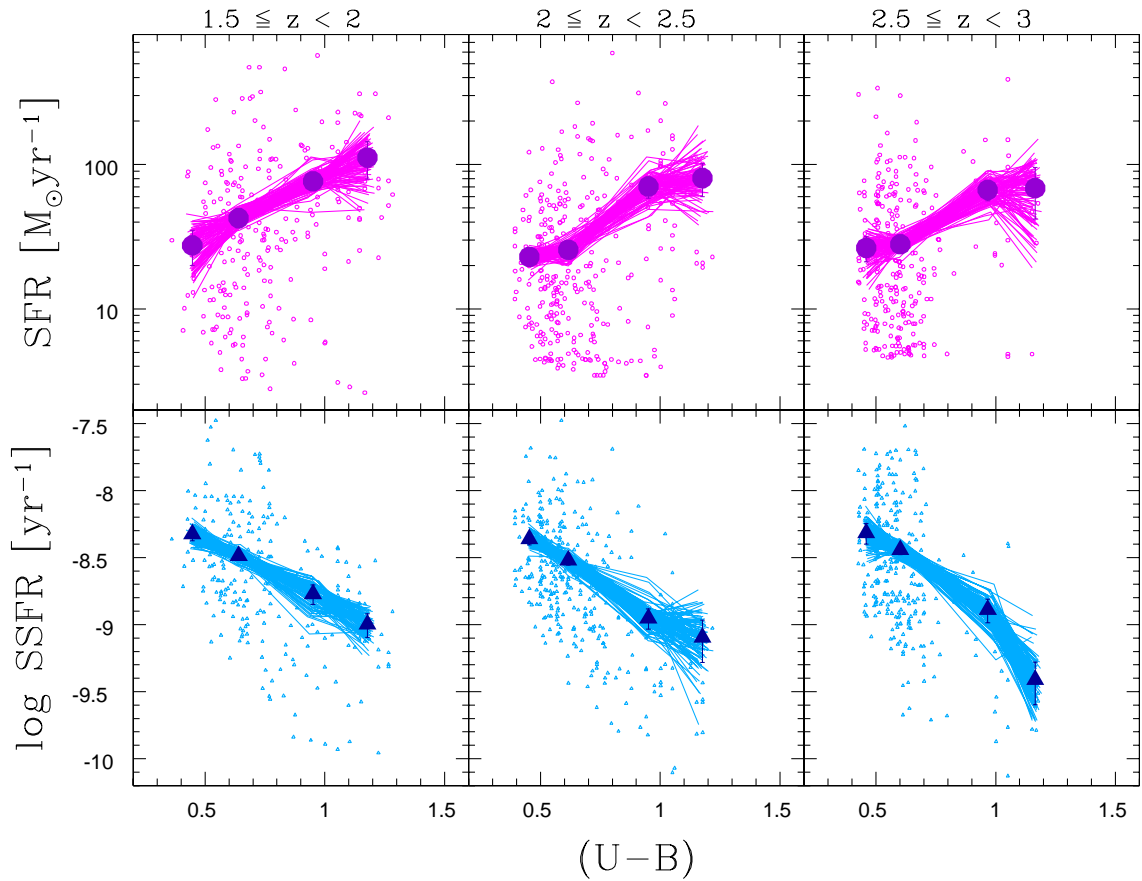}
\caption{Star formation rate (SFR, top) and specific star formation rate (SSFR, bottom) as a function of rest-frame colour $(U-B)$. Same symbols as in Figure~\ref{SFR_SSFR_SED}. }
\label{SFR_SSFR_colour}
\end{figure*}

The results of the Monte Carlo simulations are used throughout the paper to discuss the relations between SFRs, local density and stellar mass in the following sections, and are plotted in the respective figures. This allows us to demonstrate the uncertainties introduced by the photometric redshift error and to asses the reliability of the observed trends. In all figures the original data points are plotted as small symbols, whereas each Monte Carlo run is represented by a solid line showing the average in bins of local density (figures~\ref{SFR_SSFR_SED} and \ref{SFR_dens}) or stellar mass (figure~\ref{SFR_mass}) or colour (figure~\ref{SFR_SSFR_colour}), as further described below. The average of all runs in each bin is shown in the figures as big solid symbols with corresponding errorbars. We use the bi-weight estimator for location and scale, following the definition of \citet{Bee90}, to obtain a robust estimate of the average and scatter in each bin. This method was originally devised for measuring the velocity dispersion of galaxy clusters, but can be generally applied to measure a robust mean and dispersion in a population following a non-Gaussian distribution with significant fractions of outliers as well as small sample sizes.

\section{Results}\label{Results}

In the following we investigate the relations between star formation rates (SFR), specific SFR (SSFR) and local density. The specific star formation rate is defined as the SFR per unit of stellar mass, $M_\ast$, i.e., SSFR = SFR / $M_\ast$. It gives the relative importance of star formation with respect to the already existing stellar mass of a galaxy. In the following sections we will first present the SFR-density relation in three redshift bins between $z=1.5$ and $z=3$ over the whole stellar mass complete sample down to $\log~M_\ast = 9.5$ (Section~\ref{The SFR-density relation}). We then investigate the connection to the colour-density relation found in the same sample by \citet{Gru11b} and the effect of dust extinction (Section~\ref{colour and dust}) as well as the role of stellar mass in the SFR-density relation (Section~\ref{role of mstar}). For this purpose we split the sample in low and high quartile of stellar mass and local density respectively. Finally, in Section~\ref{30kpc} we show the influence of the very local environment (number of neighbours within 30~kpc) on star formation rates and rest-frame colours.

\subsection{The SFR-density relation}\label{The SFR-density relation}

Figure~\ref{SFR_SSFR_SED} shows SFR (top) and SSFR (bottom) of all individual galaxies in our sample down to the completeness limit of $\log~M_\ast = 9.5$ as a function of relative local density. The observed data points are shown as small symbols. No strong trend between SFR or SSFR and local density is present in our data. 

To asses the significance of the result we also plot the results of the Monte Carlo simulations. Each solid line in Figure~\ref{SFR_SSFR_SED} represents one Monte Carlo run, averaged in bins of local density of $\Delta \log~(1+\delta_3) = 0.6$. The average SFR and SSFR of each bin is plotted against the average $\log~(1+\delta_3)$ for each run. The average and scatter over all runs is shown as big magenta circles (SFR) and blue triangles (SSFR) in the respective panel.
The three columns show the results in three redshift bins. To show possible evolution in the above relations, we have divided our redshift range of $1.5\leq z \leq 3$ into three bins with a bin size of $\Delta z = 0.5$. The first column then corresponds to the redshift range of $1.5\leq z \leq 2$, the second to $2\leq z \leq 2.5$ and the last to $2.5\leq z \leq 3$.

The top panel shows the SFR-density relation in the three redshift bins. We do not see a strong general trend for increasing or decreasing SFR with local density at any redshift, apart from a lower average SFR in the highest local density bin at the lowest redshift ($1.5\leq z \leq 2$). The difference in average SFR between over-densities and average/under-dense areas is about a factor of three. A K-S test shows that it has a significance of just over $2 \sigma$. The average specific star formation rates (SSFRs) show a similar behaviour with local density as the average SFRs. There is a decrease of SSFR with local density in the lowest redshift bin ($1.5\leq z \leq 2$) with a significance of $\sim2 \sigma$. 

For comparison reasons we have plotted the local SFR-density relation from \citet{Lew02} and the SSFR-density relation at $z\sim0.6-0.9$ from \citet{Pat11}. Both data-sets are converted into relative over-densities by normalizing the absolute densities by the median density of the whole sample quoted in each study. We have then scaled both, the median SFRs and the median SSFRs by a factor of 10 to keep a convenient scale in the figure. Both data-sets are plotted as red squares in the panel of the first redshift bin in Figure~\ref{SFR_SSFR_SED}.
The slope of both, the SFR- and SSFR-density relation is considerable steeper in the lower redshift samples. This will be further discussed in section~\ref{summary}.

We do not find any correlation between SFR or SSFR and local density at $z>2$. This might partly be due to the increasing scatter between the single Monte Carlo simulations. Due to the scatter at low and high densities we cannot rule out the presence of a change in average SFR of a factor of up to three (at the $2\sigma$ level) between the extremes of local density. Larger samples will be needed to address this in more detail.

\subsection{The correlation to restframe $(U-B)$ colour and the effect of dust extinction}\label{colour and dust}

Galaxy colour is often used as a proxy for star-formation activity and history, however heavy dust extinction can conceal substantial amounts of current star formation leading to red rest-frame colours indicative of an old stellar population \citep[e.g.,][]{Bau11}. Our data allows us to assess if the extrapolation from colours to star-formation activity is valid at $z>1.5$ and how the two properties correlate with local environment.

So how does the SFR-density relation compare to the colour-density relation in the same data-set?
\citet{Gru11b} find that overall there is no strong correlation between $(U-B)$ colour and local density at $z > 1.5$, however, the highest over-densities tend to be inhabited by a higher fraction of blue galaxies than intermediate and low density environments. If red colour implies less star formation activity, then this trend seems to be opposite to what we find here for the SFRs and SSFRs, which both tend to be lower in the highest over-densities, however only in the lowest redshift range we probe here ($1.5\leq z \leq 2$).
This apparently contradictory behaviour of colours and SFRs in the highest density bin can be explained by the positive correlation between SFR and colour at $z>1.5$: the redder the galaxies the higher their average SFR. This trend is illustrated in Figure~\ref{SFR_SSFR_colour}, which shows the average SFR and SSFR as a function of $(U-B)$ colour. The SFRs of red galaxies are boosted by the dust-correction which strongly correlates with rest-frame $(U-B)$ colour.
This trend is discussed in more detail by \citet{Bau11}. They show that the extinction values increase from $A_V \sim 1$ at $(U-B) \sim 0.5$ to $A_V > 3$ at $(U-B) > 1$.  This dependence is supported by both the UV-slope $\beta$ and the 24$\micron$-detection fraction, which both increase with $(U-B)$ colour.  

The specific star formation rates decrease with colour, as expected, since both colour and SSFR trace the relative importance of the current star formation with respect to the total stellar population of a galaxy. This means that the high SFRs of red galaxies are due to the on average higher stellar masses of red galaxies (see also the discussion in \citet{Gru11b}).



The dust-correction is a major source of uncertainty in the determination of SFRs from UV light, as discussed in \citet{Bau11}. The dust-correction might bias the SFRs of red galaxies towards higher values, in case their red UV colour is due to an evolved stellar population rather than dusty star formation. 
This possible bias would naturally introduce a positive correlation between colour and SFR in the sample, such that red galaxies have higher SFRs. Any existing trend between colour and local density would then result in a similar relation between SFR and local density.  
Note, however, that the trend of SFR and SSFR decreasing with local density is present in both dust-corrected and uncorrected SFRs at $1.5<z<2$. As stated above, although the dust correction derived from the UV-slope has a large  uncertainty, it agrees well with the 24$\micron$-detection fraction, a completely independent indicator of dust-content \citep{Bau11}.


\begin{figure}
\includegraphics[width=0.45\textwidth]{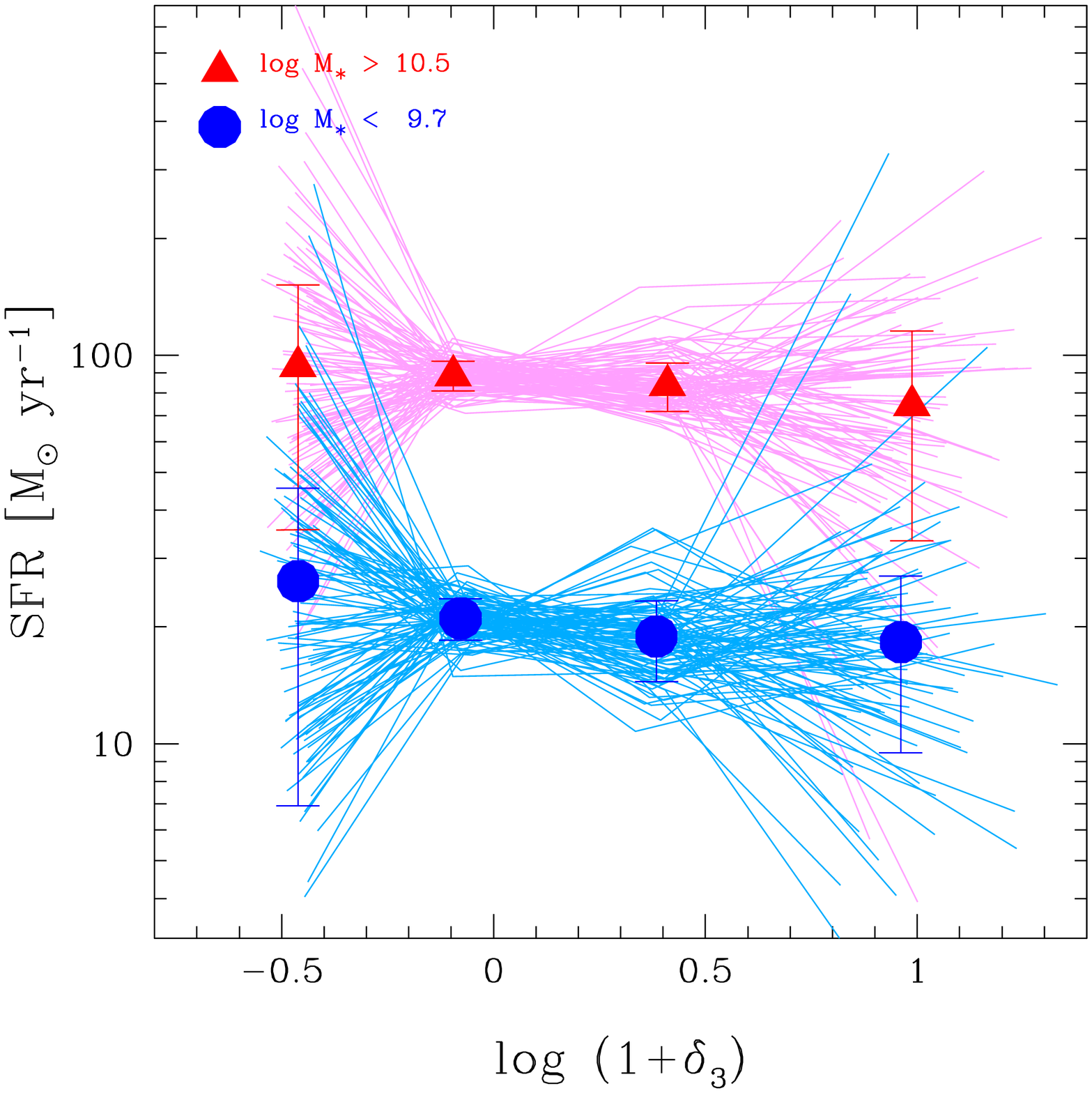}
\includegraphics[width=0.45\textwidth]{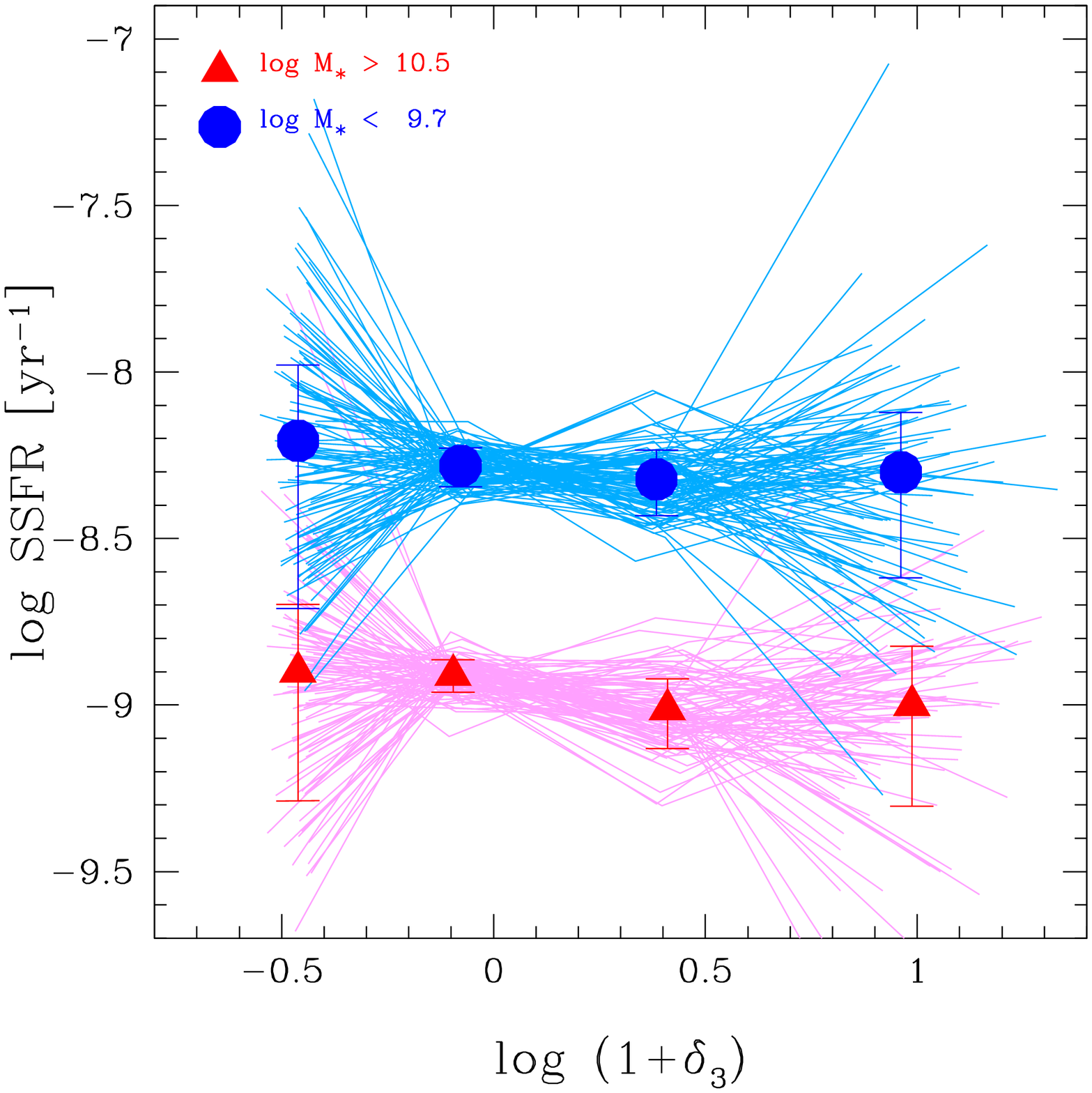}
\caption{Monte Carlo simulation of SFRs (top) and SSFRs (bottom) as a function of local density $\log ~(1+\delta_3)$ in the high (red, triangles) and low (blue, squares) quartile of stellar mass. }
\label{SFR_dens}
\end{figure}

\begin{figure}
\includegraphics[width=0.45\textwidth]{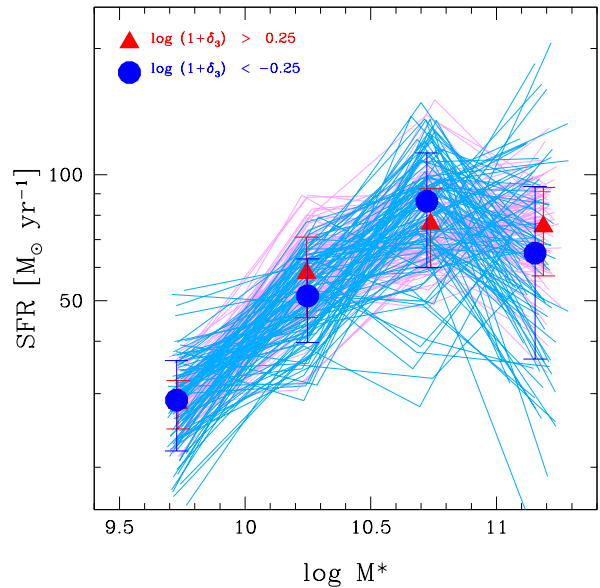}
\includegraphics[width=0.45\textwidth]{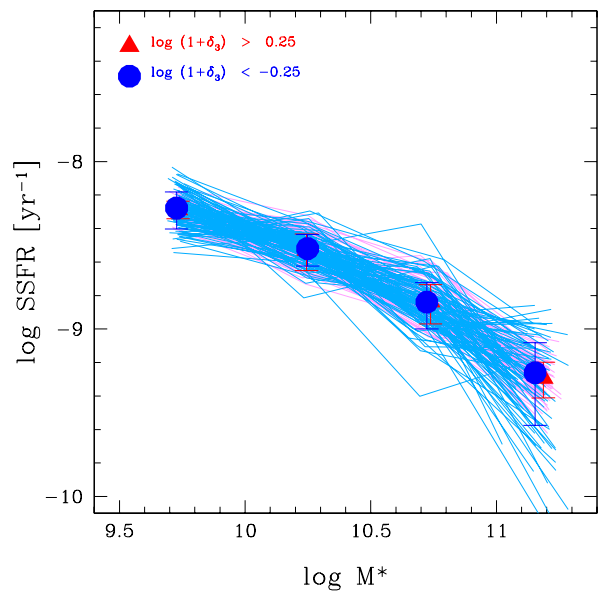}
\caption{Monte Carlo simulation of SFRs (top) and SSFRs (bottom) as a function of stellar mass $\log~M_\ast$ in the high (red, triangles) and low (blue, squares) quartile of local density. }
\label{SFR_mass}
\end{figure}

\subsection{The role of stellar mass in the SFR-density relation}\label{role of mstar}

It has been suggested by various authors that the dependence of galaxy properties on local density varies strongly with galaxy stellar mass \citep{Cas07,Bol09,Tas09,Pan09,Iov10,Gru11a,Sob10}. 
To investigate the impact of stellar mass on the SFR-density relation in our sample we now look at the SFR-density relation at ``fixed'' stellar mass. For comparison reasons we also investigate the relation between SFR and stellar mass at fixed local density. For this purpose we split the sample in high and low quartiles of stellar mass as well as high and low quartiles of local density. Note that due to our limited sample size we show the results for the whole redshift range $1.5 \leq z \leq 3$ in these figures. To avoid confusion we plot only the Monte-Carlo-simulated data and average and dispersion in each bin of local density or stellar mass, as described in the beginning of Section~\ref{Results}. 

First we investigate the behaviour of average SFR and SSFR with density in the high and low quartile of the stellar mass distribution. The sample is split in the low quartile at $\log~M_\ast < 9.7$ and in the high quartile at $\log~M_\ast > 10.5$, containing by definition 25\% of the sample respectively.  Figure~\ref{SFR_dens} shows the SFR-density (top panel) and SSFR-density (bottom panel) relation in high (red, solid triangles) and low (blue, solid squares) quartile of stellar mass. We find no trend of SFR or SSFR with density within the errors, either for low mass or high mass galaxies. 

Figure~\ref{SFR_mass} shows the average SFR and SSFR as a function of stellar mass. The colour coding is the same as before, but now the sample is split in high and low quartiles of local density, including galaxies at $\log~(1+\delta_3) > 0.25$ and $\log~(1+\delta_3) < 0.25$ in the high and low quartile respectively. There is a strong correlation between SFR and stellar mass at all densities. The average SFR increases monotonically with stellar mass until about $\log~M_\ast \sim 11$, where the relation flattens out at an average value of $\langle$SFR$\rangle \sim 100$ M$_\odot$ yr$^{-1}$. The same can be seen in the relation between SSFR and stellar mass in the bottom panel of Figure~\ref{SFR_mass}. There is a strong anti-correlation between SSFR and stellar mass (as also discussed by \citet{Bau11}), analogous to the strong correlation between colour and stellar mass in the same sample discussed in \citet{Gru11b}. The correlation steepens towards the high mass end, as a result of the flattening of the SFR-stellar mass relation discussed above. This suggests that the relative importance of star formation in the most massive galaxies is already declining at very early times ($z\sim3$), possibly connected to AGN-feedback. 

To investigate this possibility we compare the SFRs of likely AGN candidates in our sample selected via X-ray and IR-excess by Weinzirl (2011, submitted) with galaxies of similar stellar mass that show no signs of hosting an AGN. We do not see any trend for a different distribution of SFRs in AGN hosts at fixed stellar mass, neither is the flattening of the SFR-mass relation affected by the presence of AGNs in our sample. This is in agreement with a study of X-ray selected AGN host galaxies at $0.4<z<1.4$ by \citet{Bun08}, who find that in the reddest and most massive host galaxies, AGN may not be directly responsible for quenching star formation. 
The flattening could then simply be caused by the quick exhaustion of the galaxies' cold gas supply following a period of an extreme star-burst in the most massive galaxies at even higher redshifts that we probe here ($z>3$), possibly connected to galaxy merging \citep{Con08}.

\begin{figure*}
\includegraphics[angle=270, width=0.9\textwidth]{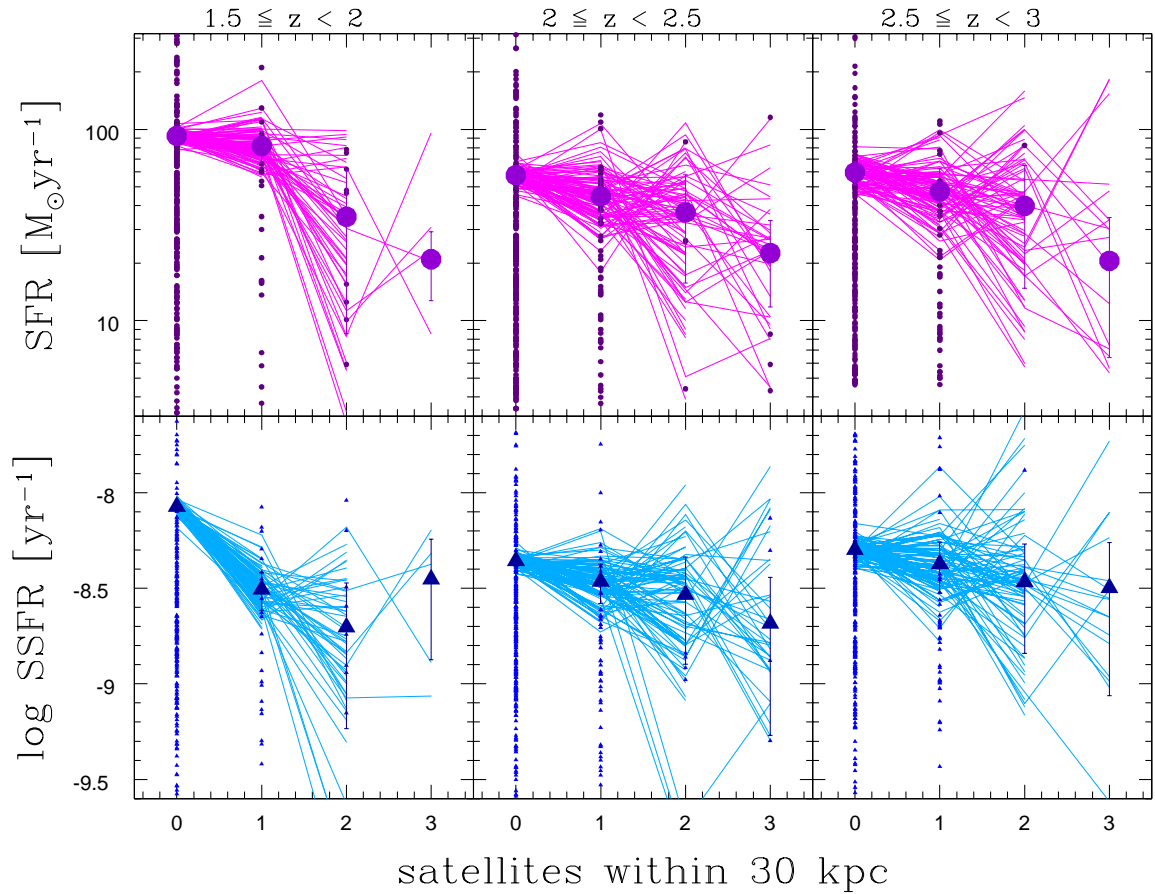}
\caption{SFR and SSFR as a function of number of close neighbours. The original data is shown as small symbols, whereas the results of the Monte Carlo simulations is overplotted as lines (one for each individual simulation) and big symbols (overall average). }
\label{SFR_30kpc}
\end{figure*}

The large scatter between the individual simulations at the low and especially the high density end is visible in all figures. This scatter is caused by the low number of objects in extreme densities. The scatter is also much larger at high stellar masses, due to the intrinsic scarcity of the most massive galaxies. The large scatter at high stellar masses above $\log~M_\ast>10.5$ could hide a difference in SFRs between the density extremes of a factor of $\sim5$ at the $2\sigma$ level. In the low mass quartile the dispersion is much smaller, but still possibly hiding a factor of $\sim2.5$. We conclude that the present sample is not well suited to investigate the behaviour of the SFR-density relation at different stellar masses due to the limited cosmological volume of the survey, which is especially problematic for high mass galaxies. However, we can exclude the presence of a strong SFR-density relation comparable to the local relation for galaxies below $\log~M_\ast=10.5$.

\begin{figure*}
\includegraphics[angle=270, width=0.9\textwidth]{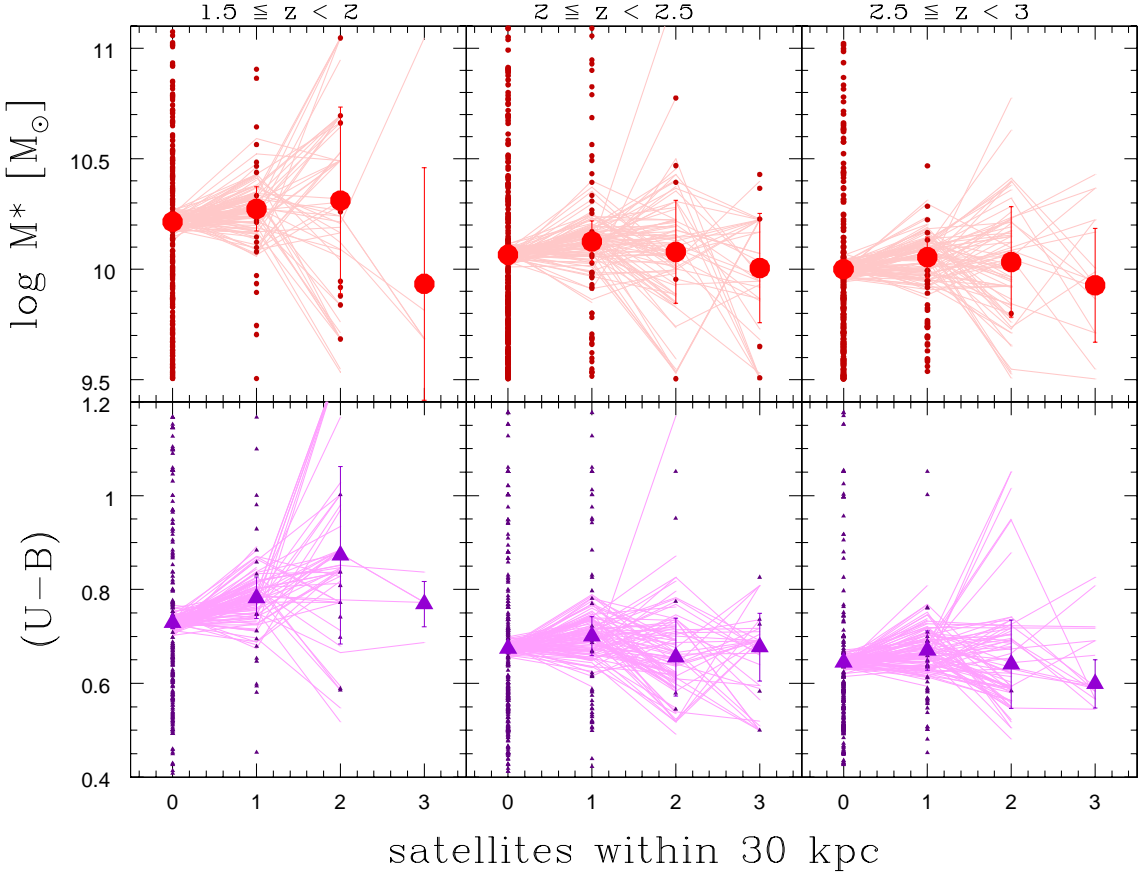}
\caption{Stellar mass and rest-frame colour as a function of number of close neighbours. The original data is shown as small symbols, whereas the results of the Monte Carlo simulations is overplotted as lines (one for each individual simulation) and big symbols (overall average). }
\label{Mast_UB_30kpc}
\end{figure*}

\subsection{The role of the very local environment within 30 kpc}\label{30kpc}

Which mechanism might be mainly responsible for the on average lower SFRs at the highest over-densities observed in our sample at $z<2$? One possible origin would be the earlier and faster formation of more massive galaxies in higher density environments \citep[e.g.,][]{Tho05}. This would imply that there should be a similar relation between local density and stellar mass, such that high-mass galaxies tend to be located in high over-densities. Another possible mechanism could be galaxy merging, which can trigger a short-lived burst of star formation exhausting a galaxy's gas reservoir and subsequently suppressing star formation, however, the time-scales involved in this process are very uncertain. \citet{Con08} have shown that the galaxy merger rate peaks at $z\sim2$, using galaxy morphologies and asymmetries in the light distribution (CAS parameters, \citet{Con03}). \citet{Blu09} using the massive galaxies from our sample ($\log~M_\ast > 11$) have found that close neighbour counts within a radius of 30~kpc are a good tracer of the galaxy merger rate up to $z\sim3$, leading to very similar results as the morphology based merger rates in \citet{Con09}. Furthermore they find that the peak in merger fraction happens earlier for massive galaxies compared to less massive ones. This enhanced merging activity could then induce a higher star-formation activity in massive galaxies dependent on redshift and on the number of close neighbours.

To test the above hypotheses we now investigate the behaviour of SFR and SSFR, stellar mass and rest-frame colour as a function of the number of neighbours within a radius of 30~kpc.
For this purpose we count the number of galaxies located within a 30 kpc co-moving radius and a certain redshift interval for each galaxy in our sample. To minimise foreground and background contamination the same redshift interval of $\Delta z = \pm 0.25$ as in the local density computation is used here.

Figure~\ref{SFR_30kpc} shows the SFR and SSFR as a function of number of neighbours within 30~kpc. The original data are plotted as small symbols (discrete values), whereas the Monte Carlo simulated data is shown as solid lines for each individual randomisation and as big symbols for the average and spread of all runs. Interestingly, we see the same trend of lower average SFR at higher local densities as for $\log~(1+\delta_3)$, but here the SFR is decreasing more gradually with number of neighbours. The trend is most visible in the lower redshift bin and weakens with increasing redshift. Notice that there are only very few objects with three or more neighbours.
We estimate the significance of this trend through a K-S test between the SFRs and SSFRs of galaxies with two or more neighbours and those of galaxies with less than two neighbours. This suggests that the difference in average SFR is significant at the 3$\sigma$ level in the lowest redshift bin ($1.5\leq z \leq 2$) and at $\sim$2$\sigma$ up to $z \sim 2.5$. The average SSFR of galaxies with two or more neighbours are significantly lower (3$\sigma$) than the SSFRs of galaxies with less than two neighbours up to $z\leq 2.5$. 

Is this trend due to a variation of stellar mass or colour? Figure~\ref{Mast_UB_30kpc} shows the dependence of stellar mass and rest-frame colour on the number of close neighbours. We see no significant dependence of average stellar mass or $(U-B)$ on the number of close neighbours. The lower average SFR for galaxies with close neighbours then seems to be not a stellar mass induced trend but is genuinely connected to a galaxy's environment.

\section{Discussion}\label{summary}

The major conclusion we draw from this study is the presence of a strong dependence of SFR and specific SFR on galaxy stellar mass, which is already detectable at $z\sim3$ \citep[see also ][]{Bau11}. The average SFR increases with stellar mass reaching values of SFR $\sim100$ at $\log~M_\ast \sim 11$. We find a flattening of the SFR-stellar mass relation at the highest stellar mass end ($\log~M_\ast > 11$), possibly connected to AGN feedback. However, we do not see significantly different SFRs or specific SFRs in AGN candidates at fixed stellar mass. The average specific SFR decreases with stellar mass, indicating that star formation activity in massive galaxies is already beginning to decline at $z\sim 3$.

In contrast to the strong correlation between SFR and stellar mass, overall we do not see a strong correlation between SFR and local density, as it is seen in the local universe \citep[e.g.,][]{Lew02} and up to $z \sim 1$ \citep[e.g.,][]{Pat09}. However, we find evidence for a possible decrease in average SFR, as well as SFR per unit mass (SSFR), at the highest local densities at $1.5\leq z\leq 2$. Galaxies in over-densities of more than a factor of five relative to the average local density exhibit on average lower SFRs as well as lower SSFRs than galaxies in average and under-dense regions. The difference amounts to a factor of 2-3 which is less than the difference in the SFR-density relation in the local Universe \citep{Lew02,Bal04} and up to $z\sim1$ \citep{Pat09,Pat11}, who find a difference of at least a factor of five over a density range of two orders of magnitude. We do not detect any differences in the SFR-density relation if the sample is split in low and high quartiles of stellar mass, while the SFR-stellar mass relation is equally strong at both high and low local densities.

Our data is also compatible with no correlation between SFR and local density, as was also found in a study of Lyman Break Galaxies (LBGs) at $z\sim3$ by \citet{BL05}. Note, however, that a sample of LBGs naturally includes only blue, star-forming galaxies. As the authors point out, such a sample might not allow for the detection of a SFR-density relation present in a mass-complete sample. Similarly, the SFRs for star-forming galaxies at $z\sim0$ do not vary strongly with local density \cite[see e.g.,][]{Bal04}.

Compared to the intermediate redshift range of $z\sim1-2$ our findings are not consistent with the increasing SFRs with density in field galaxies reported by e.g., \citet{Elb07} and \citet{Coo08}. Higher star-formation in the most dense environments was also found in high redshift galaxy clusters at $z=1.46$ \citep{Hay10} and $z=1.62$ \citep{Tra10}. Other studies of a massive galaxy cluster at $z\sim1.4$ however show the opposite trend of quenched star formation \citep{Bau10} and older stellar populations \citep{Lid08} of galaxies in the cluster centre. Similar results were found recently by \citet{Chu11} and \citet{Qua11} studying the colour-density and SFR-density relations at $z\sim 1-2$ in more general, field dominated samples. Both studies use the UKIDSS (UKIRT Infrared Deep Sky Survey) Ultra Deep Survey (UDS) and slightly different but similar techniques of measuring local densities (fixed aperture and nearest neighbour densities). The above authors show that galaxies in higher local densities have on average redder colours \citep{Chu11} and lower star-formation rates \citep{Qua11} out to $z\sim1.8$, even at fixed stellar mass. This is consistent with the redshift range in which we find on average slightly lower SFRs in the most over-dense environments in this study.

One issue in comparing galaxy properties with local environment might be the different ranges of absolute local densities probed by different studies, as pointed out by \citet{Sob10}. They use narrow band H$\alpha$ imaging over a wide area to show that the median SFRs increase with local densities in environments that correspond to the field and into cluster outskirts, but drop sharply towards cluster core densities. Depending on the range of absolute densities and total mass of the structures that are probed by a given study, the result could vary from increasing over constant to decreasing SFRs with local density. To estimate the large scale density or total mass range of structures we are likely to probe in our sample we apply the cluster abundance function found by \citet{Bah03} to our survey volume. The abundances of structures of different mass is most likely evolving with cosmic time, however, it is not well constrained from observations at high redshift. We therefore use the local cluster abundances of \citet{Bah03} to get a rough estimate. We find that we can expect only a few structures of a total mass within $0.6 h^{-1}$Mpc of $M_{0.6} > 10^{14} M_\odot$ to be present in our sample. This suggests that in this study we are confined to the field and group environment and that it is very unlikely that we are probing environments with the highest absolute densities like massive galaxy clusters in this study. According to the results of \citet{Sob10} we should then find on average increasing SFRs with local density, which we do not see in our sample.

We also investigate the influence of the presence of close neighbours within a radius of 30~kpc on star formation rates and find that galaxies with more close neighbours show lower average SFRs as well as lower SSFRs up to $z<2.5$. This is consistent with what is found for the $3^{rd}$ nearest neighbour densities above, but the relation between SFR and number of close neighbours has a higher significance than the SFR-density relation. We also see a more gradual decline of the average SFR with number of neighbours than with local density. The relation between SFR and number of close neighbours is not caused by variations in average stellar-mass. Star formation quenching induced by the very local environment in the highest density regions, possible due to galaxy merging, seems to be present in our sample at least up to $z\sim2$ and possibly up to $z\sim2.5$. 

Another possible explanation for our results is that local density traces the dark matter halo mass of the large scale structure in which a galaxy resides. The halo mass is expected to determine the time-scale and amount of cold gas accreted by a galaxy, as predicted by the hydrodynamical simulations of e.g., \citet{Ker05}, who model the evolution of cold gas accretion and star formation rates over cosmic time. Since the cold accretion fraction depends on local density - due to typically higher halo masses in high density areas - they also predict decreasing SFRs with local density. The typical density at which the SFR starts to decrease progressively shifts to higher densities at higher redshift and is detectable up to $z\sim2$, but not at $z\sim3$. This prediction is consistent with the low average SFR at the highest densities we see at $z\sim1.5-2$.

\section{Conclusions}

We study the relationship between star formation, local density and stellar mass at a redshift of $1.5 \leq z \leq 3$ using data from a very deep, near-infrared HST-survey, the GOODS NICMOS Survey (GNS). We find the following three main conclusions:

\begin{enumerate}
\renewcommand{\theenumi}{\arabic{enumi}.}

\item Star formation rates strongly depend on galaxy stellar mass over the whole redshift range of $1.5<z<3$ we study here. The correlation between SFR as well as specific SFR and stellar mass does not depend strongly on environment and is similar at all local densities.

\item The influence of the local environment on star formation rates at $z>1.5$ is much weaker than in the local universe (and up to $z\sim1$). The average SFR does not decrease monotonically with density, but is suppressed only in the highest density regions, and only up to $z\sim2$, where over-densities of a factor of $>5$ have on average lower SFRs (by a factor of $\sim3$) than average and under-dense areas. This is in good agreement with expectations from hydrodynamical simulations \citep{Ker05}, which predict that lower SFRs at the highest over-densities should be detectable up to $z\sim2$.

\item The presence of close neighbours within 30~kpc has a similar effect on the average SFRs as high 3$^{rd}$ nearest neighbour densities. Overall both local density indicators show the same trend, but the decrease of average  SFR in galaxies with several close companions is more significant than the decrease of average SFR at high relative over-densities. This suggests that the very local environment has an influence on the star formation activity in early galaxies possibly through the mechanism of galaxy merging.

\end{enumerate}

\noindent In conclusion our data supports the emerging picture where stellar mass largely drives galaxy evolution. Local environment seems to have little impact on the star formation rates of galaxies at $z>1.5$, apart from the most dense environments, possibly the most massive structures, which already had time to build up and exert their influence on galaxies after a lifetime of the Universe of only $\sim$3 Gyrs.

\section*{Acknowledgments}

We would like to thank Jes\'us Varela and Carlos Hoyos for many useful discussions and very helpful suggestions. We acknowledge funding from the UK Science and Technology Facilities Council (STFC). CJC acknowledges financial support from the Leverhulme Trust. We also thank the referee for their constructive report and useful comments.

\end{document}